\documentclass{article}

\PassOptionsToPackage{square,numbers}{natbib}
\usepackage[preprint]{neurips_2024}
\usepackage{booktabs}
\usepackage{float}

\usepackage[utf8]{inputenc} 
\usepackage[T1]{fontenc}    
\usepackage{hyperref}       
\usepackage{url}            
\usepackage{booktabs}       
\usepackage{amsfonts}       
\usepackage{nicefrac}       
\usepackage{microtype}      
\usepackage{xcolor}         
\usepackage[pdftex]{graphicx}
\usepackage{multirow}
\usepackage{amsmath}

\title{Unlocking Telemetry Potential: Self-Supervised Learning for Continuous Clinical Electrocardiogram Monitoring}

\author{
\hspace{-0.9cm} \textbf{Thomas Kite PhD}$^{1,2}\thanks{Corresponding author}$ \quad \textbf{Uzair Tahamid Siam}$^{1}$ \quad \textbf{Brian Ayers MD MBA}$^{3}$ \\
\hspace{-0.9cm} \textbf{Nicholas Houstis MD PhD}$^{1,4}$ \quad \textbf{Aaron D Aguirre MD PhD}$^{1,2,4,5}$ \\
\hspace{-0.9cm} $^1$Cardiology Division, Massachusetts General Hospital, Harvard Medical School; Boston, MA, U.S.A.\\
\hspace{-0.9cm} $^2$Wellman Center for Photomedicine, Massachusetts General Hospital; Boston, MA, U.S.A.\\
\hspace{-0.9cm} $^3$Cardiac Surgery Division, Massachusetts General Hospital, Harvard Medical School; Boston, MA, U.S.A.\\
\hspace{-0.9cm} $^4$Healthcare Transformation Lab, Massachusetts General Hospital; Boston, MA, U.S.A.\\
\hspace{-0.9cm} $^5$Center for Systems Biology, Massachusetts General Hospital; Boston, MA, U.S.A.\\
\hspace{-0.9cm} \texttt{\{tkite,usiam,bayers,nhoustis,aaguirre1\}@mgh.harvard.edu}\\
}

\begin{document}

\maketitle

\begin{abstract}
Machine learning (ML) applied to routine patient monitoring within intensive care units (ICUs) has the potential to improve care by providing clinicians with novel insights into each patient's health and expected response to interventions. This paper applies deep learning to a large volume of unlabeled electrocardiogram (ECG) telemetry signals, which are commonly used for continuous patient monitoring in hospitals but have important differences from the standard, single time-point 12-lead ECG used in many prior machine learning studies. We applied self-supervised learning to pretrain a spectrum of deep networks on approximately 147,000 hours of ECG telemetry data. Our approach leverages this dataset to train models that significantly improve performance on four distinct downstream tasks compared with direct supervised learning using labeled data. These pretrained models enable medically useful predictions and estimates in smaller patient cohorts that are typically limited by the scarcity of labels. Notably, we demonstrate that our pretrained networks can continuously annotate ECG telemetry signals, thereby providing monitoring capabilities that are often unavailable due to the requirement for specialized expertise and time-consuming professional annotations.
\end{abstract}

\section{Introduction}
\label{sec:introduction}
The modern intensive care unit (ICU) is a high-stakes time-pressured environment in which physicians must process a large volume of complex data to make decisions. Continuous patient monitoring, often of the cardiovascular system, exemplifies this data processing challenge. Such data are often in the form of continuous electrical waveforms recording every beat of the heart for days at high sampling frequencies. Unaided, a physician can at best extract a tiny fraction of the health information embedded in these measurements. The ability of modern machine learning (ML) algorithms to process large amounts of diverse data could enable automated extraction of insights from continuous monitoring, providing novel metrics for patient progress and offering early warning signs for preventive intervention.

An increasingly common application of ML to cardiovascular health is analysis of the electrocardiogram (ECG). This intricate electrical signal originates from the coordinated activity of all the cells in the heart, which can be measured on the skin surface as electrical potentials. Encoded in this electrical activity is a wealth of information about a patient's health and as a result ECG is often continuously monitored in hospitalized patients. The application of convolutional neural networks (CNNs) \cite{LeCun:1989:cnn_original} to ECG analysis has yielded many surprising and promising results \cite{Hannun:2019:ambulatory_afib, Zachi:2019:ecg_resnet_afib_original, Ribeiro:2020:ecg_resnet}, uncovering nuanced patterns in the signal beyond human perception with important implications for a patient's condition.

Most applications of CNNs to the ECG have focused on analyzing the \textit{gold-standard} 12-lead ECG, a standardized measurement obtained at a point in time, but processing continuous ECG telemetry remains an open challenge. A 12-lead ECG measurement is a 10-second recording, made and vetted by a human operator, resulting in high signal quality at the cost of the operator's time. The resource-intensive nature of the measurement limits its use, often as a reactive measurement to a change in the clinical status of a hospitalized patient. In contrast, ECG telemetry is a continuous measurement and requires minimal human oversight, enabling applications such as the prediction of future clinical events and the monitoring of patient trajectories. To date, automated analysis of continuous ECG telemetry has yet to be fully explored using modern ML techniques. Extending previous works on 12-lead ECG to telemetry signals presents both difficulties and opportunities. For example, the telemetry signal can be plagued with artefacts from human motion, instrumental noise, missing packets in the data stream, pacemaker signals, and many other sources of distraction for neural networks (see appendix~\ref{app:ecg_visualisation} for examples). Furthermore, there are known differences in the signal filtering used for telemetry compared to the 12-lead ECG signal \cite{Venkatachalam:2011:ecg_filtering}, which could change the morphology of ECG waveforms in important ways. Solving these challenges, however, would enable important new decision support opportunities, as the quantity of ECG telemetry data for a given patient is orders of magnitude larger than the data from 12-lead ECGs. Additionally, automated telemetry analysis could provide a longitudinal view of a patient's health by studying uninterrupted hours of data instead of intermittent 10 second segments.

One pervasive challenge in clinical prediction applications is the limited availability of high quality target labels. This limitation stems from the desire to predict rare events, analyze progressively smaller subsets of patients, or to predict complex endpoints that require adjudication from highly trained professionals. Label sparsity creates significant challenges for training deep learning models on high dimensional data such as ECG telemetry. 

In this paper we use self-supervised learning (SSL) applied to a large archive containing millions of hours of ECG telemetry to address label sparsity in clinical prediction tasks. We benchmark these pretrained models on multiple downstream prediction tasks, verifying that larger models are more capable of ingesting the deluge of data, with notable gains compared to supervised learning from scratch. We first demonstrate this pretraining advantage on three large \textit{proxy} tasks which evaluate the model's performance across different label quantities, and subsequently on a medically useful downstream task with naturally scarce labels. Additionally we present illustrations of how the fine-tuned models can be used to continuously annotate the telemetry signal with clinically-useful metrics of the patient's cardiac electrical activity, a technique which promises to drive new insights into pathophysiology and therapeutic effects.

\section{Related work}
\label{sec:related_work}
\paragraph{Supervised learning with CNN and ECG.} The ECG is a multivariate periodic time-series of electrical potentials. CNNs have emerged as one of the most popular architectures for analyzing such signals given their capacity to detect patterns in waveform shape, amplitude, and duration. An early and noteworthy example is \cite{Hannun:2019:ambulatory_afib} which classified 12-lead ECG rhythms with impressive accuracy -- surpassing expert human cardiologists. Another early and surprising result came from \cite{Zachi:2019:ecg_age_sex}, in which a CNN was used to predict the age and biological sex of patients, information which was not previously known to be embedded in the ECG. Many works have built on this foundation to predict aspects of blood chemistry \cite{Galloway:2019:hyperkalemia_dl}, predict future onset of atrial fibrillation \cite{Zachi:2019:ecg_resnet_afib_original,Gadaleta:2023:afib_cnn_lstm, Kolk:2024:vae_ecg}, and to measure time intervals from the 12-lead ECG \cite{Alam:2024:qt_interval}.

\paragraph{Self-Supervised Learning for ECG.} While the integration of CNNs in ECG analysis has proven highly effective, it also presents several challenges, including the need for large labeled datasets. To address applications with label scarcity, researchers have explored self-supervised learning methods that leverage large volumes of unlabeled 12-lead ECG data. Work by \citet{Kiyasseh:2020:clocs_original} demonstrated a method that uses signal transformations, similar to those introduced by \citet{Chen:2020:contrastive_learning_images} for images, as well as semantic information from consecutive signals and distinct lead positions, to provide multiple perspectives of the same patient. Building on this foundation, \citet{Diamant:2022:pclr} employed SSL by using signals from the same patient, even those recorded years apart, as a basis for positive pairing. This patient contrastive approach yielded impressive results on downstream prediction tasks. More recently, \citet{Abbaspourazad:2023:foundation_wearable} used similar SSL approaches for analyzing single-lead ECG data collected from wearable devices. For a broader perspective on the advancement of SSL in domains beyond computer vision, see the comprehensive review by \citet{Deldari:2022:beyond_vision_pretraining_review}. 

\section{Dataset description}
\label{sec:dataset_description}
The dataset for this work was recorded from telemetry monitoring systems at 
Massachusetts General Hospital (MGH),
containing high frequency waveforms and vital signs from 150k unique patients collected between 2014 and 2023.\footnote{This work was performed under approval by the Institutional Review Board at the MGH (Protocol 2020P003053).
}
In particular we focused on ECG telemetry derived from four leads (generally leads I, II, III and V1) sampled at 120Hz, representing >17M total hours of ECG telemetry data.
From the bulk data we curated a high-quality dataset for model training. For every patient, we divided their hospital stay into one-hour blocks. Within each block we removed ECG values in excess of $\pm 60$mV. We then created continuous uninterrupted 60 second segments from the remaining data and selected the least noisy segment as measured by the squared sum of low (<0.75Hz) and high (>40Hz) frequency Fourier modes, which was chosen empirically in early testing. These steps yielded a final dataset containing 8.85M ECG segments from 72k unique patients.

The four downstream tasks in this work come from three main sources. First, we combine the bulk ECG dataset with the dates of birth and biological sex for all patients, thus giving two tasks: \textbf{age regression} and \textbf{sex classification}. This allows a comparison to common downstream tasks in the literature \cite{Zachi:2019:ecg_age_sex, Diamant:2022:pclr, Abbaspourazad:2023:foundation_wearable}.
Second, we employ a novel strategy of extrapolating labels from the annotated 12-lead ECG, which occur routinely alongside the persistent monitoring for some patients. By synchronising the two signal sources we have extrapolated the high quality professional annotations to 90k 60s telemetry waveforms from 19k patients. In particular for this paper we use the annotated timing intervals to perform \textbf{intervals regression}. The four intervals are QRS duration, QT interval, PR interval and ventricular rate (i.e. heart rate). See appendix~\ref{app:ecg_visualisation} for an illustration of these intervals.
Third, we curate a cohort of patients who have undergone thoracic surgery and are known to be at risk for developing post-operative atrial fibrillation (Afib), a cardiac arrhythmia caused by chaotic electrical activity of the top chambers (atria) of the heart that can predispose to stroke \cite{Aggarwal:2023:postop_afib}. Of the patients that develop Afib we select the first 10 minutes in that state, and for the rest we simply take the first 10 minutes of telemetry signal available (both divided into 60s segments, subject to the same $\pm 60$mV filtering criteria mentioned above). In total this dataset contains 1551 patients and 14.4k 60 second ECG segments which we use for \textbf{Afib classification}.

\section{Training methodology}
\label{sec:training_methodology}
All results use ResNet encoders\footnote{Models made available at \href{https://github.com/TomKite57/ResNet1D}{github.com/TomKite57/ResNet1D}} adapted from \cite{He:2015:resnet_original, He:2016:resnet_v2} for multivariate 1D time-series (see appendix~\ref{app:architecture} for further details of the model architecture).
The training was split between a self-supervised pretraining stage and a downstream supervised learning stage. For the prediction stage we used either a linear evaluation with frozen weights or a fine-tuning approach. For reasons discussed in Sect.~\ref{sec:method_ssl_for_ecg}, the dataset was split at the patient level rather than the ECG segment level. A 90\%--10\% split between training and validation was used consistently across pretraining and downstream tasks. All computation was performed on local compute for data privacy reasons with all models trained on a single NVIDIA 4090 GPU with 24GB of VRAM. As such GPU memory was a potentially limiting factor in training which we mitigated by using the momentum based memory mechanism proposed by \citet{He:2019:moco_original}.
Pretraining typically occupied 2--4 hours depending on the model size, while downstream training is much faster at 2--40 minutes depending on model size and label quantity. Without additional experiments the results in this paper could be reproduced in 3--4 days on a single 4090 GPU, however the total computing time is closer to two weeks given early testing/tuning and experiments not discussed here.
All models and training were implemented using PyTorch (version 2.2.1) \cite{paszke:2019:pytorch}. Pretraining additionally made use of PyTorch Lightning (version 2.2.3) \cite{falcon:2019:pytorch_lightning}.

\subsection{Self-supervised learning for ECG}
\label{sec:method_ssl_for_ecg}
The pretraining stage used a contrastive based pretext learning task. Following the work of \citet{Chen:2020:simCLR_V2} a 3-layer MLP was attached to the encoder following the global average pool and subsequently discarded for downstream prediction tasks. In order to cope with arbitrary model sizes with limited GPU VRAM we employed a momentum based approach \cite{He:2019:moco_original}. Below the main conceptual choices are discussed, and hyperparameter values are provided in appendix~\ref{app:hyperparameters} (Table~\ref{tab:pretraining-hyperparameters}).

To pretrain using contrastive learning, we first define a positive-pairing of ECG segments as those recorded from the same individual, regardless of the date \cite{Diamant:2022:pclr}. The findings of both \cite{Diamant:2022:pclr, Abbaspourazad:2023:foundation_wearable} suggest that the semantic information available by associating individuals is more powerful than associating time and space \cite{Kiyasseh:2020:clocs_original} or from using data augmentations such as signal masking, simulated instrument noise and baseline deviation noise \cite{Abbaspourazad:2023:foundation_wearable, Kiyasseh:2020:clocs_original, Mehari:2021:ssl_for_12_lead}. This patient contrastive learning of representations (PCLR) approach yielded impressive results on both the pretext task (as measured by patient identifiability \cite{Diamant:2022:pclr}) and subsequent downstream tasks. Given that patients are identifiable, we mitigated the potential for data leakage by dividing training and validation sets at the patient level rather than the ECG level. Furthermore we defined an epoch as a single randomly selected ECG segment from each patient, thereby preventing the model from memorizing patients with more available hours of ECG, while also creating an inherent randomness to each epoch that could act as a regularising effect. There were typically $\sim100$ available 60s ECG segments for each patient.

As first described in \cite{He:2019:moco_original} we used momentum based constrative learning (MoCo), which involves maintaining two encoders: the query encoder which is updated using backpropagation and a key encoder which is updated using a momentum-based moving average of the query encoder. Similar to \cite{He:2019:moco_original} we maintained a $K$ length fixed size queue of data samples: the encoded representations of the current mini-batch are enqueued, and the oldest are dequeued. In our case, a query and key within the minibatch were a match if the two ECG were from the same patient and everything else in the queue was considered a negative pairing. We then used InfoNCE loss between query $q$ and positive key $k_+$ at temperature $\tau$,
\begin{equation}
\label{eq:info_nce}
    \mathcal{L}_{\text{InfoNCE}} = -\log \frac{\exp(q \cdot k_+ / \tau)}{\sum_{i=0}^K \exp(q \cdot k_i / \tau)},
\end{equation}
and averaged over the minibatch of positive pairs (negative pairs are only present in the queue, not the minibatch). This loss function maximizes the mutual information between the positive pair representations. In addition to using InfoNCE loss for optimization we also tracked the closely related NT-Xent loss \cite{Chen:2020:contrastive_learning_images} between positive pairs $z_i$ and $z_j$ with all other embeddings within the minibatch serving as negative pairs,
\begin{equation}
\label{eq:nt_xent}
    \mathcal{L}_{\text{NT-Xent}} = -\log \frac{\exp(\text{sim}(z_i, z_j) / \tau)}{\sum_{k=1}^{2N} \mathbf{1}_{[k \neq i]} \exp(\text{sim}(z_i, z_k) / \tau)},
\end{equation}
where ${\rm sim}$ denotes Cosine similarity. This loss function was also averaged over a minibatch, though the denominator makes Eq.~\eqref{eq:nt_xent} still batch-size dependent, so we always evaluate on a minibatch of 512 ECG segments. We tracked this quantity to allow easier comparison to previous results \cite{Diamant:2022:pclr} and future comparisons to this work.

The MoCo approach allowed the total exposure to negative samples to be decoupled from the batch size without incurring the expected performance penalty of a small minibatch \cite{Chen:2020:simCLR_V2}. For the purposes of this work, limiting the batch size enabled us to more easily scale the model with a consistent set of hyperparameters and thus produce reliable scaling relationships for model size (see Fig.~\ref{fig:proxy_downstream_percent_gain}).

\subsection{Downstream prediction: linear-evaluation and fine-tuning}
To apply the pretrained encoders to a downstream prediction task we first discarded the MLP projection head and then attached a new linear head which was first trained on the task with all other parameters frozen. In the case of fine-tuning this was followed by updating all parameters simultaneously with a slow learning rate ($\sim 10^{-4}$). Our objective was not to exhaustively optimize every model on every downstream task, but rather to conduct a comparative analysis to illuminate the contributions of model size, label scarcity, and pretraining on task performance. As such, we adopted a strategy in which a baseline set of hyperparameters are kept as uniform as possible. We only altered the total epochs, batch sizes and learning rates when needed for computational feasibility or numerical stability. Additionally we applied no explicit regularization, using only the implicit regularization entailed by randomness during data loading (see Sect.~\ref{sec:method_ssl_for_ecg}) and finite batch sizes \cite{Smith:2021:implicit_regularization_sgd}. Though some models presented in Sect.~\ref{sec:results} could thus be further optimized with regularization and hyperparameter turning, such improvements would certainly become insignificant for sufficiently small label quantities, and as a result we would not expect the scaling relations reported in Sect.~\ref{sec:results} to materially change. Fiducial hyperparameters and notable exceptions are reported in appendix~\ref{app:hyperparameters} (Table~\ref{table:downstream_hyperparameters}).

The loss functions we used for model fitting included mean square error (MSE) for age regression, binary cross entropy loss for sex/Afib classification, and mean absolute error (MAE) normalised by the standard deviation of each label distribution for the intervals task. We selected the final model using performance metrics of MAE for age, area under the receiver operating characteristic (AUROC, henceforth AUC for brevity) for sex/Afib and mean absolute percentage error (MAPE) for intervals. Note that these loss functions correlate well with other metrics used in the literature (e.g. $F1$, $R^2$), but slightly higher performing models could in principle be selected for these (see further discussion in appendix~\ref{app:study_comparison}).

\section{Results}
\label{sec:results}
In order to probe the quality of the pretrained models we evaluated each downstream task on a variety of model sizes \cite{Chen:2020:simCLR_V2}. In anticipation of those results we first note that the figures of merit achieved for each task are within the ranges expected from other works, keeping in mind however that comparison between works without shared datasets is nuanced for both signal quality and demographic reasons. For example, one point of comparison for this work is \citet{Abbaspourazad:2023:foundation_wearable} in which the participants are Apple watch owners -- markedly younger and healthier than the cohort we study which contains hospitalized patients. It was noted in \cite{Zachi:2019:ecg_age_sex} that a CNN-predicted age above a patient's actual age was indicative of cardiac disease, which is prevalent in hospitalized cohorts. The main figures of merit for each task and demographic characteristics of relevant studies are summarised in appendix~\ref{app:study_comparison}.

\subsection{Pretraining with patient contrastive loss}
\begin{figure}
\centering
\includegraphics[width=1.0\textwidth]{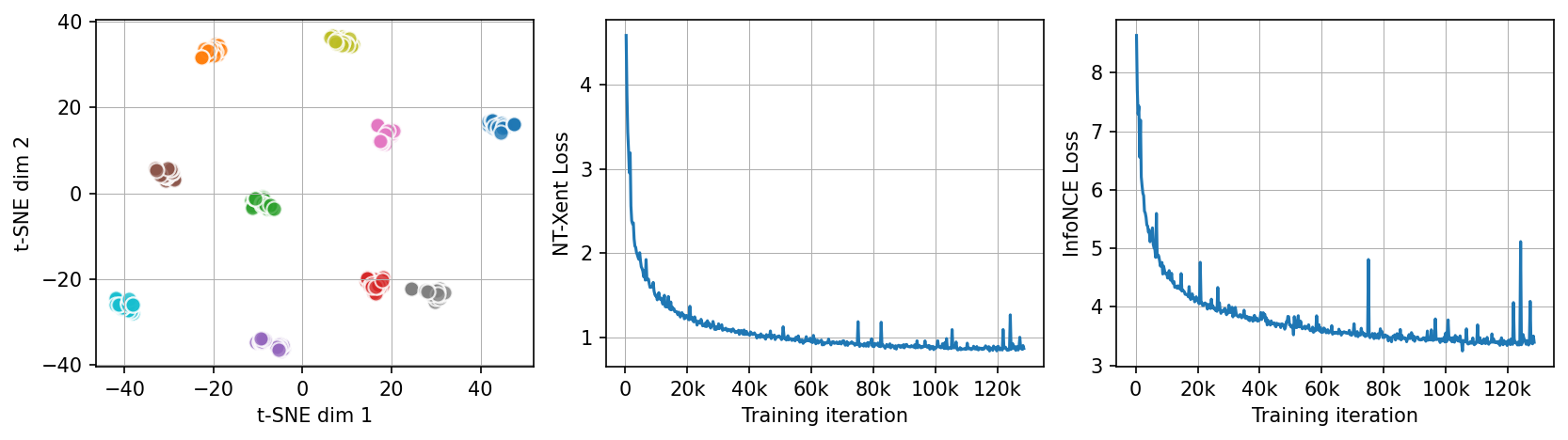}
\caption{Pretraining using patient-contrastive learning applied to ECG telemetry. (Left) t-SNE plot illustrating the distribution of embeddings corresponding to different participants, highlighting distinct and compact clusters. (Center) NT-Xent loss and (Right) InfoNCE loss across training iterations, demonstrating rapid initial decrease followed by stabilization.}
\label{fig:pretraining-plot}
\end{figure}
Results from the pretraining stage are shown in Fig.~\ref{fig:pretraining-plot}. In the left-most panel it can be seen that the embeddings of 200 different ECGs arising from 10 patients are indeed all tightly clustered in the 2D t-SNE visualisation. Importantly these are patients from the validation set, so the learned representation induces tight clustering even on patients not seen during training. The center and right-most panel show the NT-Xent and InfoNCE losses respectively for a characteristic model. Though performance in the pretext task does not necessarily correlate with performance on the downstream tasks \cite{Chen:2020:contrastive_learning_images}, these loss curves indicate stable learning dynamics and can be useful for comparisons across studies.

\subsection{Larger models benefit from pretraining}
\label{sect:results_model_scale}
\begin{figure}
\centering
\includegraphics[width=1.0\textwidth]{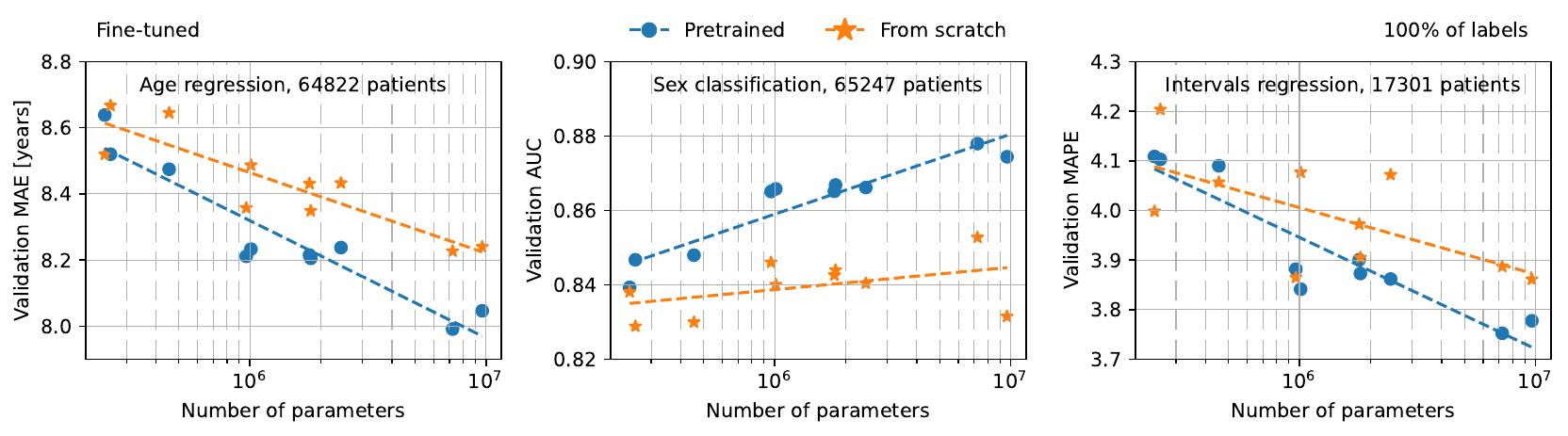}
\caption{Pretrained and subsequently fine-tuned models outperform models trained from scratch. Model performance is plotted as a function of model size and pretraining (vs from scratch) across three prediction tasks: age regression (left), sex classification (center) and intervals regression (right). Models were trained with 100\% of available labels. Trend lines emphasize scaling relations.}
\label{fig:proxy_downstream_absolute_100}
\end{figure}
\begin{figure}
\centering
\includegraphics[width=1.0\textwidth]{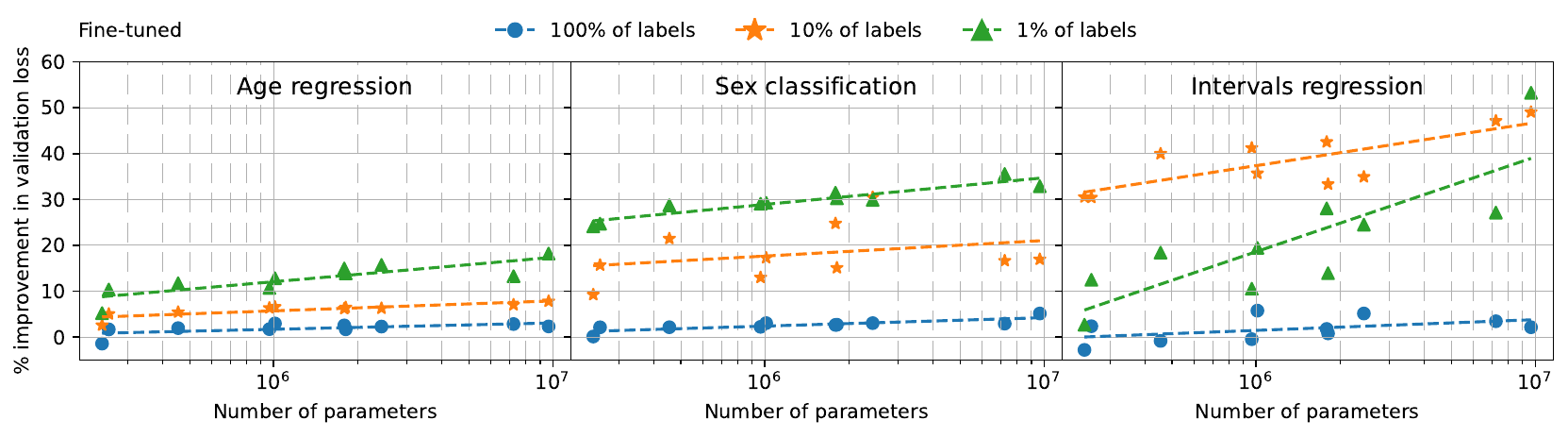}
\caption{The advantage of pretrained networks grows with model size and scarcity of labels. Performance advantage is measured as \% improvement in validation loss compared to models trained from-scratch on the same subset of labels.}
\label{fig:proxy_downstream_percent_gain}
\end{figure}
The age, sex and intervals downstream tasks (introduced in Sect.~\ref{sec:dataset_description}) all benefit from the availability of abundant high-quality labels, making them ideal for testing the impact of the pretraining stage. We first evaluated the impact of pretraining on each prediction task using 100\% of the downstream labels across a spectrum of model sizes. We found that fine-tuned models (pretrained) outperformed from-scratch models (no pretraining) across virtually all model sizes. Moreover the performance advantage reliably grew with the size of the model (Fig.~\ref{fig:proxy_downstream_absolute_100}).

The advantage of pretraining was further magnified as labels became more scarce (Fig.~\ref{fig:proxy_downstream_percent_gain}), mirroring the performance gains seen with self-supervised learning in image classification tasks \cite{Chen:2020:simCLR_V2}. Pretrained models that were fine-tuned with 10\% or 1\% of labels improved their performance by up to 20-40\% compared to from-scratch models trained on the same subset of labels. In the intervals regression task we saw a distinct performance response to label scarcity. Pretrained models still performed better than from-scratch models as labels became scarce, but the improvement was more pronounced with 10\% of labels than 1\% of labels. This may reflect an inductive bias imposed on the ECG latent space by the patient based pretraining task \cite[See][for discussion of inductive bias within SSL]{Bendidi:2023:ssl_no_free_lunch}. Encouraging ECGs from the same patient to have similar embeddings despite hours, days or even years of difference, explicitly instructs the model to ignore features of the ECG which fluctuate with time while also being generic to many patients, e.g. heart rate. Meanwhile age and sex are effectively constants for many of the patients studied here. If there is an inductive bias against encoding intervals like heart rate then some number of labels would be needed to break that bias for downstream tasks, a penalty not incurred by purely supervised models. Further evidence for this inductive bias can be seen in appendix~\ref{app:absolute_downstream} (Fig.~\ref{fig:proxy_downstream_linear_eval_absolute_grid}), where we showed that linear evaluation without fine-tuning can still perform well on the age and sex tasks, but fails to surpass any supervised intervals model. This hints that the final ECG latent space learned by PCLR may not itself be very versatile, but the intermediate layers have useful weights and biases that are easily tuned to unlock the structure learned during the pretraining stage.

\subsection{Application to post-operative atrial fibrillation monitoring}
\label{sect:poaf_monitoring}
\begin{figure}
\centering
\includegraphics[width=1.0\textwidth]{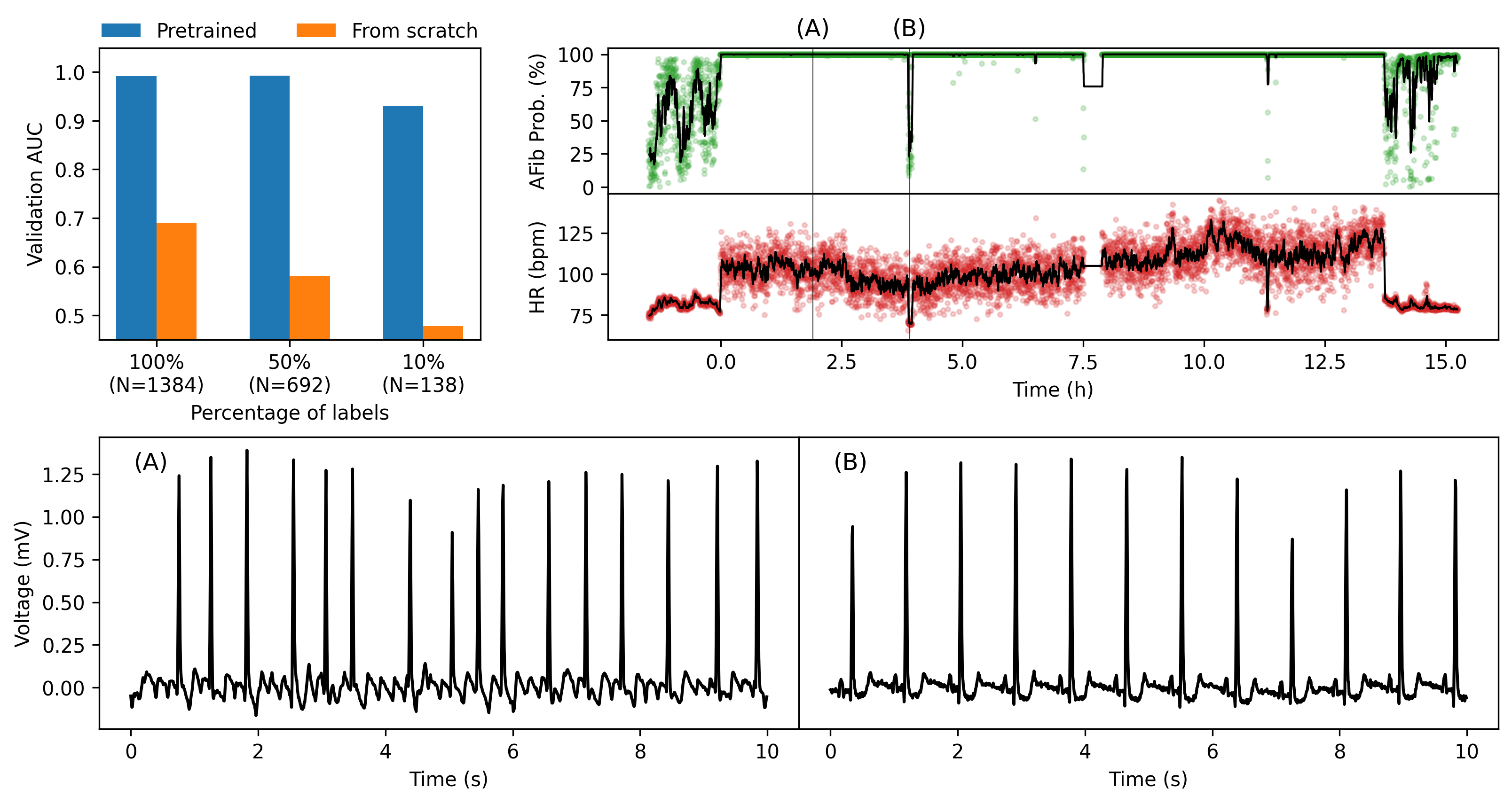}
\caption{Classification of atrial fibrillation from continuous ECG telemetry signals using a pretrained neural network. (Top left) Afib classifier performance as a function of label scarcity. (Top right) Probability of Afib (top) estimated across a continuous 15-hour stretch of ECG telemetry together with HR prediction (bottom). Both the Afib probability and heart rate reveal the onset of the cardiac arrhythmia. (Lower panels) the corresponding ECG segments from two moments of interest in the longer time series. The leftmost ECG shows an arrhythmia (Afib), while the segment on the right is a small window where the heart regained normal sinus rhythm.}
\label{fig:main_afib_figure}
\end{figure}
The Afib classification task represents a clinically meaningful study with a true scarcity of labels, providing a real-world test of the utility of pretrained models. The post-operative cohort is of key interest since a relatively large number of these patients ($16$ -- $50\%$) will develop post-operative Afib \cite{Aggarwal:2023:postop_afib}, and consequently they are often monitored continuously with ECG telemetry during their hospitalization. In this case the scarcity of labels arises from seeking an overlap of conditions: following a specific surgical procedure and developing a specific pathophysiology. Such a retrospective analysis is typical of clinical research, where patients are selected to satisfy a multiplicity of criteria, such as drugs taken, demographics, overlapping signal sources, prior disease states, etc. Afib and other cardiac arrhythmias have been the focus of many other studies and public datasets \cite{Hannun:2019:ambulatory_afib, Zachi:2019:ecg_resnet_afib_original, Goldberger:2000:physionet, Gadaleta:2023:afib_cnn_lstm, Diamant:2022:pclr}. In this work we chose a thoracic surgery cohort instead of the larger available datasets due firstly to its suitability to the ultimate telemetry monitoring goal, and secondly in order to target a more concrete pathway to the final physiological arrhythmic state. Post-operative atrial fibrillation in thoracic surgery patients can be driven by inflammation, direct manipulation of the heart, fluid-volume shifts or the stress of surgery. In other scenarios, long-term structural remodeling of the heart can lead to Afib, as suggested by the work of \citet{Zachi:2019:ecg_resnet_afib_original} demonstrating the ability to predict Afib more than a month prior to its first recorded occurrence.

For the detection of Afib, pretrained models again outperformed from-scratch models, but with a much larger advantage than was seen in the other tasks (Fig.~\ref{fig:main_afib_figure}). This likely stems from the much sparser set of labels, accentuating the advantage pretrained models enjoy in ECG telemetry tasks. Model size played a much lesser role here, so performance measures were averaged over model size. In the top right panel of Fig.~\ref{fig:main_afib_figure} we illustrate the potential for real time insights that could be derived from ECG-telemetry prediction tools such as the Afib classifier described here. Analyzing a 15-hour stretch of ECG telemetry in a patient from the validation set, we noted that at $t=0$ (Afib onset) there was a sudden change in heart rate and with it the Afib probability increased to $100\%$ (heart rate dysregulation is a hallmark of Afib). Furthermore we classified small segments within the Afib window as having reverted to the normal rhythm, a common occurrence for post-operative patients \cite{Aggarwal:2023:postop_afib}. These transitions were validated by visually inspecting the telemetry signal (Fig.~\ref{fig:main_afib_figure}, bottom panels A and B). Interestingly, we noted that the Afib score was noticeably positive before and after the Afib window, which rarely occurred for patients that never developed Afib. This suggests that the neural network was picking up on subtle changes in the ECG prior to overt Afib, changes that may be undetectable to a human, thereby corroborating the recently described ability of CNNs to predict Afib as well as detect it \cite{Zachi:2019:ecg_resnet_afib_original, Gadaleta:2023:afib_cnn_lstm}.

\subsection{Monitoring cardiac effects of drug administration}
\label{sect:pharmacokinetics_monitoring}
\begin{figure}
\centering
\includegraphics[width=1.0\textwidth]{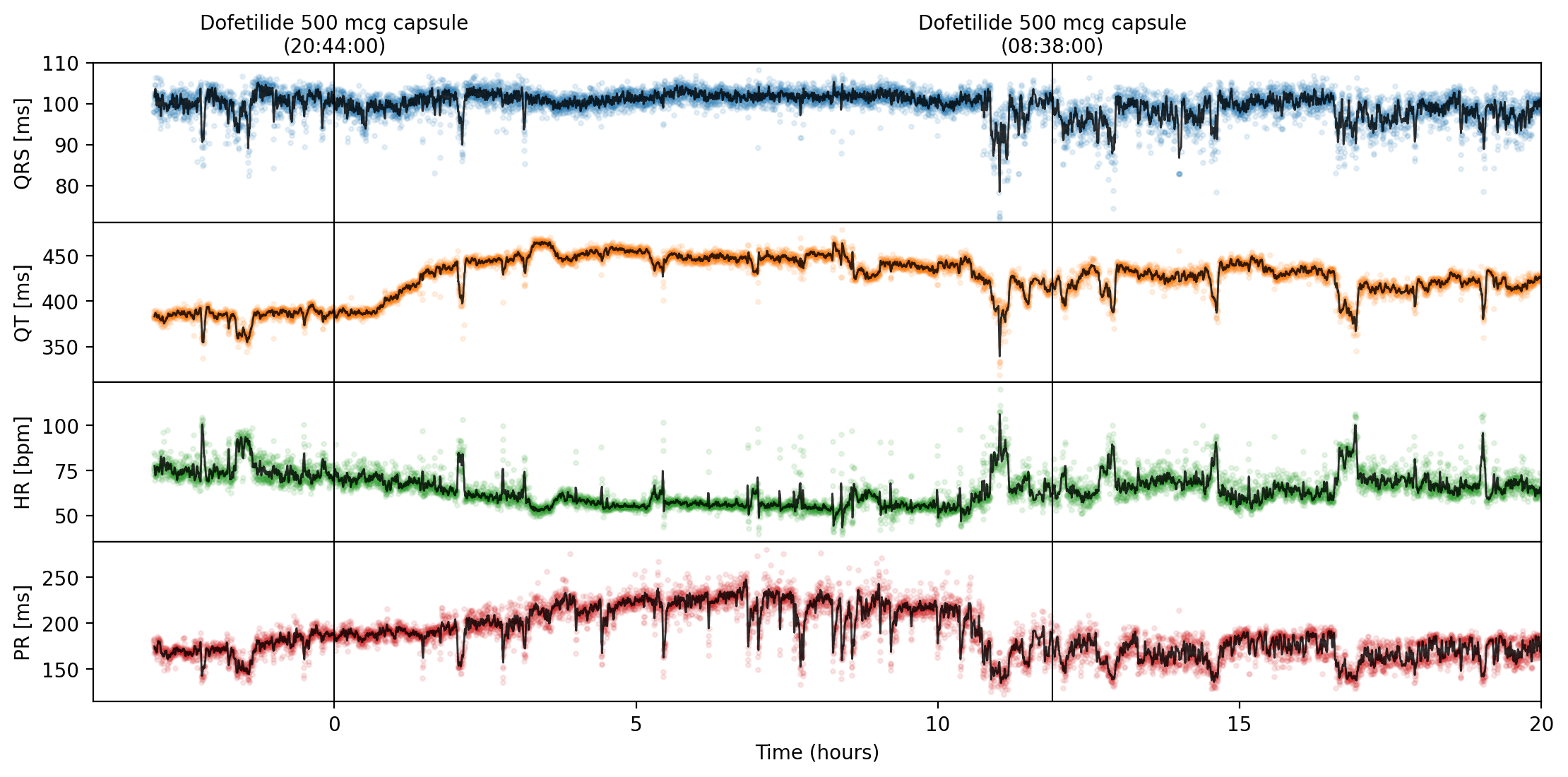}
\caption{ECG timing intervals reliably change after a first dose of dofetilide (500 mcg capsule), as measured using a neural network designed for continuous telemetry signals. Both QT and PR intervals lengthen in the hours following the dose.}
\label{fig:dofe_night}
\end{figure}
As a further demonstration of how the fine-tuned models can annotate telemetry to aid decision making we show a case study of patients receiving a high-risk medication. Dofetilide, a drug which inhibits specific electrical signaling in the heart, is administered to prevent certain cardiac arrhythmias but can itself also cause new, potentially lethal arrhythmias if not monitored carefully. In Fig.~\ref{fig:dofe_night} we used our intervals regression model to estimate the effects of dofetilide on the cardiac intervals of a patient during the initial 20 hours following the first dose. The neural network detected prolongation of both the QT and PR intervals, with a corresponding decrease in heart rate (see appendix~\ref{app:ecg_visualisation} for depiction of timing intervals). These changes were confirmed clinically with a 12-lead ECG. A QT interval that exceeds 500ms increases risk of developing a dangerous ventricular arrhythmia, highlighting the value of such tools for continuous intervals monitoring in the clinical setting. \citet{Alam:2024:qt_interval} similarly predict QT intervals using a single ECG lead, and as a case study show a patient whose QT interval spiked to 500ms following a dose of dofetilide, which in turn triggered an ICU alarm. This work expands on that same concept by enabling a continuous measurement based on telemetry ECG.

Interval monitoring is a use-case that illustrates a broader strategy in which a pretrained model can be fine-tuned on sparse high quality labels and subsequently used to annotate millions of hours of retrospective ECG telemetry. Doing so in this case elucidated a characteristic response to a single drug, but the same principle could be applied to the study of nontrivial interaction of multiple drugs. This same approach could be repeated with non ECG-labels that are similarly resource-intensive and therefore sparse, such as invasive hemodynamics or cardiac ejection fraction.

\section{Discussion and conclusion}
\label{sec:discussion_and_conclusion}
This paper aims to establish a robust foundation of powerful models and techniques capable of analyzing the vast amount of data collected in continuous ECG telemetry patient monitoring. The strategy demonstrated in this paper uses self-supervised learning to effectively ingest the vast data available in this domain. The consequent fine-tuned models outperformed the supervised models on every task considered, and it was shown that for very small label amounts the gains were far more pronounced. Using these models for inference on real-world patient data showed how they can effectively extrapolate sparse high quality labels to annotate the millions of hours of available ECG telemetry. Using these techniques for real time monitoring carries enormous potential for one day affecting clinical decision making and the standard of care more generally.

Due to the enormous success of CNNs applied to ECGs from the last five years \cite{Zachi:2019:ecg_resnet_afib_original, Hannun:2019:ambulatory_afib, Galloway:2019:hyperkalemia_dl, Diamant:2022:pclr} we focused on the potential for scaling ResNets to the task of telemetry signal processing. In doing so we showed that large models benefit enormously from the pretraining stage, enabling medical tasks which would otherwise be inaccessible (see Fig.~\ref{fig:main_afib_figure}). While it is possible that with hyperparameter tuning, parameter sweeps and strong regularization some of the supervised models we trained from scratch could have performed better, our results strongly suggest there is some label quantity for which large pretrained models will always have the advantage. Additionally, training large models which adapt easily to multiple downstream tasks is a promising foundation for ML based ECG-telemetry interpretation. A clear future direction is to extend this to new architectures which are better suited for ingesting arbitrarily long data, spanning hours or days, considering that this is a core advantage of ECG telemetry. Early testing with CNNs encoding longer ECG segments did not yield clear downstream benefits, suggesting the need for new architectures not as thoroughly explored in the literature. Some authors have found little gain in applying \textit{off-the-shelf} transformer models compared to ResNets \cite{Abbaspourazad:2023:foundation_wearable}, while others have found advantage to combining CNN encoders with LSTM models \cite{Gadaleta:2023:afib_cnn_lstm}. Both architectures will be explored in future work.

Prior work on pretraining methodology for 12-lead ECG has shown that contrastive learning with positive pairs of patients outperforms a variety of alternatives based on temporal, spatial or augmented associations \cite{Diamant:2022:pclr, Abbaspourazad:2023:foundation_wearable}. Our early testing on ECG telemetry also showed worse performance with segment level pretraining. Despite the promising results from patient level pretraining, the PCLR approach may exhibit some inductive bias, as indicated by the outcomes in the intervals task (see Fig.~\ref{fig:proxy_downstream_percent_gain}). An alternative segment-level self-supervised learning approach could enhance scalability by allowing pretraining to expand with the volume of ECG data rather than being confined by the number of patients, potentially beating current benchmarks. Two promising directions include investigating the optimal mix of ECG augmentations and their hyperparameters, or designing a balance between inter- and intra-patient differentiation during pretraining \cite[e.g. see][]{Hugo:2021:neighborhood_contrastive_learning}.

Another significant opportunity offered by ICU telemetry monitoring is the analysis of simultaneous measurements of distinct physiology properties. For example, it is possible to monitor time-synchronized measures of heart electrical activity (ECG), blood oxygen saturation (photoplethysmography, PPG) and blood pressure (arterial blood pressure line, ABP). A latent space constructed from the combination of signals would offer a vastly more complete view of not just the heart but the entire cardiovascular system spanning both macro- and micro-circulation. A simple implementation of patient contrastive learning may not work for such a task however considering the surprising patient identifiability of ECG, a feature not found in PPG signals \cite{Abbaspourazad:2023:foundation_wearable} and likely not found in ABP either. Further explorations into the multi-signal pretraining methodology are warranted.

\section{Broader impact}
\label{sec:broader_impact}
The methods developed in this work can facilitate the effective interpretation of ECG telemetry across the large subset of hospitalized patients who are monitored by continuous ECG. Automated ECG interpretation could bring expert level analysis to care environments with fewer resources or access to specialists. More generally, the use of self-supervised learning in the medical domain amplifies the performance of models trained with scarce labels, either because the events are rare, the label adjudicators are few, or the patient subsets are highly selected. \cite[see for example broader impact statement in][]{Chen:2020:simCLR_V2}.

A major consideration for the use of ML in healthcare is interpretability of the results to clinicians. Because data interpretation drives high-stakes decision making, healthcare providers must have confidence in the model outputs, ideally through an ability to reconcile model recommendations with clinical understanding of disease pathophysiology. However, even if interpretability of a model is limited, there must be some confidence interval placed on its outputs.

A final important consideration is the possible bias imbued in models from the training data. Demographics and clinical characteristics of patients may vary between different hospitals, or even for the same hospital over time. To tackle this, diverse training data sets involving multiple institutions will be needed, and validation tests should be constructed on targeted demographic groups and in different institutions.

\bibliographystyle{plainnat}
\bibliography{lit}

\begin{thebibliography}{31}
\providecommand{\natexlab}[1]{#1}
\providecommand{\url}[1]{\texttt{#1}}
\expandafter\ifx\csname urlstyle\endcsname\relax
  \providecommand{\doi}[1]{doi: #1}\else
  \providecommand{\doi}{doi: \begingroup \urlstyle{rm}\Url}\fi

\bibitem[{Abbaspourazad} et~al.(2023){Abbaspourazad}, {Elachqar}, {Miller}, {Emrani}, {Nallasamy}, and {Shapiro}]{Abbaspourazad:2023:foundation_wearable}
Salar {Abbaspourazad}, Oussama {Elachqar}, Andrew~C. {Miller}, Saba {Emrani}, Udhyakumar {Nallasamy}, and Ian {Shapiro}.
\newblock {Large-scale Training of Foundation Models for Wearable Biosignals}.
\newblock \emph{arXiv e-prints}, art. arXiv:2312.05409, December 2023.
\newblock \doi{10.48550/arXiv.2312.05409}.

\bibitem[Aggarwal et~al.(2023)Aggarwal, Siems, Potel, Hingtgen, Wang, Nijjar, Huddleston, John, Kelly, and Voeller]{Aggarwal:2023:postop_afib}
R.~Aggarwal, C.~Siems, K.~N. Potel, A.~Hingtgen, Q.~Wang, P.~S. Nijjar, S.~J. Huddleston, R.~John, R.~F. Kelly, and R.~K. Voeller.
\newblock New-onset postoperative atrial fibrillation after mitral valve surgery: Determinants and the effect on survival.
\newblock \emph{JTCVS Open}, 16:\penalty0 305--320, Sep 2023.
\newblock \doi{10.1016/j.xjon.2023.08.018}.

\bibitem[{Alam} et~al.(2023){Alam}, {Aguirre}, and {Stultz}]{Alam:2024:qt_interval}
Ridwan {Alam}, Aaron {Aguirre}, and Collin {Stultz}.
\newblock {Detecting QT prolongation From a Single-lead ECG With Deep Learning}.
\newblock \emph{arXiv e-prints}, art. arXiv:2401.05378, December 2023.
\newblock \doi{10.48550/arXiv.2401.05378}.

\bibitem[Attia et~al.(2019{\natexlab{a}})Attia, Friedman, Noseworthy, Lopez-Jimenez, Ladewig, Satam, Pellikka, Munger, Asirvatham, Scott, Carter, and Kapa]{Zachi:2019:ecg_age_sex}
Zachi~I. Attia, Paul~A. Friedman, Peter~A. Noseworthy, Francisco Lopez-Jimenez, Douglas~J. Ladewig, Ganesh Satam, Patricia~A. Pellikka, Thomas~M. Munger, Samuel~J. Asirvatham, Christopher~G. Scott, Rickey~E. Carter, and Suraj Kapa.
\newblock Age and sex estimation using artificial intelligence from standard 12-lead ecgs.
\newblock \emph{Circ Arrhythm Electrophysiol}, 12\penalty0 (9):\penalty0 e007284, Sep 2019{\natexlab{a}}.
\newblock \doi{10.1161/CIRCEP.119.007284}.
\newblock Epub 2019 Aug 27.

\bibitem[Attia et~al.(2019{\natexlab{b}})Attia, Noseworthy, Lopez-Jimenez, Asirvatham, Deshmukh, Gersh, Carter, Yao, Rabinstein, Erickson, Kapa, and Friedman]{Zachi:2019:ecg_resnet_afib_original}
Zachi~I Attia, Peter~A Noseworthy, Francisco Lopez-Jimenez, Samuel~J Asirvatham, Abhishek~J Deshmukh, Bernard~J Gersh, Rickey~E Carter, Xiaoxi Yao, Alejandro~A Rabinstein, Brad~J Erickson, Suraj Kapa, and Paul~A Friedman.
\newblock An artificial intelligence-enabled ecg algorithm for the identification of patients with atrial fibrillation during sinus rhythm: a retrospective analysis of outcome prediction.
\newblock \emph{The Lancet}, 394\penalty0 (10201):\penalty0 861--867, 2024/03/02 2019{\natexlab{b}}.
\newblock \doi{10.1016/S0140-6736(19)31721-0}.
\newblock URL \url{https://doi.org/10.1016/S0140-6736(19)31721-0}.

\bibitem[{Bendidi} et~al.(2023){Bendidi}, {Bardes}, {Cohen}, {Lamiable}, {Bollot}, and {Genovesio}]{Bendidi:2023:ssl_no_free_lunch}
Ihab {Bendidi}, Adrien {Bardes}, Ethan {Cohen}, Alexis {Lamiable}, Guillaume {Bollot}, and Auguste {Genovesio}.
\newblock {No Free Lunch in Self Supervised Representation Learning}.
\newblock \emph{arXiv e-prints}, art. arXiv:2304.11718, April 2023.
\newblock \doi{10.48550/arXiv.2304.11718}.

\bibitem[{Chen} et~al.(2020{\natexlab{a}}){Chen}, {Kornblith}, {Norouzi}, and {Hinton}]{Chen:2020:contrastive_learning_images}
Ting {Chen}, Simon {Kornblith}, Mohammad {Norouzi}, and Geoffrey {Hinton}.
\newblock {A Simple Framework for Contrastive Learning of Visual Representations}.
\newblock \emph{arXiv e-prints}, art. arXiv:2002.05709, February 2020{\natexlab{a}}.
\newblock \doi{10.48550/arXiv.2002.05709}.

\bibitem[{Chen} et~al.(2020{\natexlab{b}}){Chen}, {Kornblith}, {Swersky}, {Norouzi}, and {Hinton}]{Chen:2020:simCLR_V2}
Ting {Chen}, Simon {Kornblith}, Kevin {Swersky}, Mohammad {Norouzi}, and Geoffrey {Hinton}.
\newblock {Big Self-Supervised Models are Strong Semi-Supervised Learners}.
\newblock \emph{arXiv e-prints}, art. arXiv:2006.10029, June 2020{\natexlab{b}}.
\newblock \doi{10.48550/arXiv.2006.10029}.

\bibitem[{Deldari} et~al.(2022){Deldari}, {Xue}, {Saeed}, {He}, {Smith}, and {Salim}]{Deldari:2022:beyond_vision_pretraining_review}
Shohreh {Deldari}, Hao {Xue}, Aaqib {Saeed}, Jiayuan {He}, Daniel~V. {Smith}, and Flora~D. {Salim}.
\newblock {Beyond Just Vision: A Review on Self-Supervised Representation Learning on Multimodal and Temporal Data}.
\newblock \emph{arXiv e-prints}, art. arXiv:2206.02353, June 2022.
\newblock \doi{10.48550/arXiv.2206.02353}.

\bibitem[Diamant et~al.(2022)Diamant, Reinertsen, Song, Aguirre, Stultz, and Batra]{Diamant:2022:pclr}
N.~Diamant, E.~Reinertsen, S.~Song, A.D. Aguirre, C.M. Stultz, and P.~Batra.
\newblock Patient contrastive learning: A performant, expressive, and practical approach to electrocardiogram modeling.
\newblock \emph{PLoS Computational Biology}, 18\penalty0 (2):\penalty0 e1009862, Feb 2022.
\newblock \doi{10.1371/journal.pcbi.1009862}.

\bibitem[Falcon and team(2019)]{falcon:2019:pytorch_lightning}
William Falcon and The PyTorch~Lightning team.
\newblock Pytorch lightning, March 2019.
\newblock URL \url{https://www.pytorchlightning.ai}.
\newblock If you want to cite the framework, feel free to use this (but only if you loved it).

\bibitem[Gadaleta et~al.(2023)Gadaleta, Harrington, Barnhill, Hytopoulos, Turakhia, Steinhubl, and Quer]{Gadaleta:2023:afib_cnn_lstm}
Matteo Gadaleta, Patrick Harrington, Eric Barnhill, Evangelos Hytopoulos, Mintu~P. Turakhia, Steven~R. Steinhubl, and Giorgio Quer.
\newblock Prediction of atrial fibrillation from at-home single-lead ecg signals without arrhythmias.
\newblock \emph{npj Digital Medicine}, 6\penalty0 (1):\penalty0 229, 2023.
\newblock \doi{10.1038/s41746-023-00966-w}.
\newblock URL \url{https://doi.org/10.1038/s41746-023-00966-w}.

\bibitem[Galloway et~al.(2019)Galloway, Valys, Shreibati, Treiman, Petterson, Gundotra, Albert, Attia, Carter, Asirvatham, Ackerman, Noseworthy, Dillon, and Friedman]{Galloway:2019:hyperkalemia_dl}
Conner~D. Galloway, Alexander~V. Valys, Jacqueline~B. Shreibati, Daniel~L. Treiman, Frank~L. Petterson, Vivek~P. Gundotra, David~E. Albert, Zachi~I. Attia, Rickey~E. Carter, Samuel~J. Asirvatham, Michael~J. Ackerman, Peter~A. Noseworthy, John~J. Dillon, and Paul~A. Friedman.
\newblock {Development and Validation of a Deep-Learning Model to Screen for Hyperkalemia From the Electrocardiogram}.
\newblock \emph{JAMA Cardiology}, 4\penalty0 (5):\penalty0 428--436, 05 2019.
\newblock ISSN 2380-6583.
\newblock \doi{10.1001/jamacardio.2019.0640}.
\newblock URL \url{https://doi.org/10.1001/jamacardio.2019.0640}.

\bibitem[Goldberger et~al.(2000 (June 13))Goldberger, Amaral, Glass, Hausdorff, Ivanov, Mark, Mietus, Moody, Peng, and Stanley]{Goldberger:2000:physionet}
A.~L. Goldberger, L.~A.~N. Amaral, L.~Glass, J.~M. Hausdorff, P.~Ch. Ivanov, R.~G. Mark, J.~E. Mietus, G.~B. Moody, C.-K. Peng, and H.~E. Stanley.
\newblock {PhysioBank, PhysioToolkit, and PhysioNet}: Components of a new research resource for complex physiologic signals.
\newblock \emph{Circulation}, 101\penalty0 (23):\penalty0 e215--e220, 2000 (June 13).
\newblock Circulation Electronic Pages: http://circ.ahajournals.org/content/101/23/e215.full PMID:1085218; doi: 10.1161/01.CIR.101.23.e215.

\bibitem[Hannun et~al.(2019)Hannun, Rajpurkar, Haghpanahi, Tison, Bourn, Turakhia, and Ng]{Hannun:2019:ambulatory_afib}
Awni~Y. Hannun, Pranav Rajpurkar, Masoumeh Haghpanahi, Geoffrey~H. Tison, Codie Bourn, Mintu~P. Turakhia, and Andrew~Y. Ng.
\newblock Cardiologist-level arrhythmia detection and classification in ambulatory electrocardiograms using a deep neural network.
\newblock \emph{Nature Medicine}, 25\penalty0 (1):\penalty0 65--69, 2019.
\newblock \doi{10.1038/s41591-018-0268-3}.
\newblock URL \url{https://doi.org/10.1038/s41591-018-0268-3}.

\bibitem[{He} et~al.(2015){He}, {Zhang}, {Ren}, and {Sun}]{He:2015:resnet_original}
Kaiming {He}, Xiangyu {Zhang}, Shaoqing {Ren}, and Jian {Sun}.
\newblock {Deep Residual Learning for Image Recognition}.
\newblock \emph{arXiv e-prints}, art. arXiv:1512.03385, December 2015.
\newblock \doi{10.48550/arXiv.1512.03385}.

\bibitem[{He} et~al.(2016){He}, {Zhang}, {Ren}, and {Sun}]{He:2016:resnet_v2}
Kaiming {He}, Xiangyu {Zhang}, Shaoqing {Ren}, and Jian {Sun}.
\newblock {Identity Mappings in Deep Residual Networks}.
\newblock \emph{arXiv e-prints}, art. arXiv:1603.05027, March 2016.
\newblock \doi{10.48550/arXiv.1603.05027}.

\bibitem[{He} et~al.(2019){He}, {Fan}, {Wu}, {Xie}, and {Girshick}]{He:2019:moco_original}
Kaiming {He}, Haoqi {Fan}, Yuxin {Wu}, Saining {Xie}, and Ross {Girshick}.
\newblock {Momentum Contrast for Unsupervised Visual Representation Learning}.
\newblock \emph{arXiv e-prints}, art. arXiv:1911.05722, November 2019.
\newblock \doi{10.48550/arXiv.1911.05722}.

\bibitem[He et~al.(2018)He, Zhang, Zhang, Zhang, Xie, and Li]{he:2018:bag_of_tricks}
Tong He, Zhi Zhang, Hang Zhang, Zhongyue Zhang, Junyuan Xie, and Mu~Li.
\newblock Bag of tricks for image classification with convolutional neural networks.
\newblock \emph{arXiv preprint arXiv:1812.01187}, 2018.

\bibitem[{Ioffe} and {Szegedy}(2015)]{Ioffe:2015:batchnorm_original}
Sergey {Ioffe} and Christian {Szegedy}.
\newblock {Batch Normalization: Accelerating Deep Network Training by Reducing Internal Covariate Shift}.
\newblock \emph{arXiv e-prints}, art. arXiv:1502.03167, February 2015.
\newblock \doi{10.48550/arXiv.1502.03167}.

\bibitem[{Kingma} and {Ba}(2014)]{Kingma:2014:adam_optimizer}
Diederik~P. {Kingma} and Jimmy {Ba}.
\newblock {Adam: A Method for Stochastic Optimization}.
\newblock \emph{arXiv e-prints}, art. arXiv:1412.6980, December 2014.
\newblock \doi{10.48550/arXiv.1412.6980}.

\bibitem[{Kiyasseh} et~al.(2020){Kiyasseh}, {Zhu}, and {Clifton}]{Kiyasseh:2020:clocs_original}
Dani {Kiyasseh}, Tingting {Zhu}, and David~A. {Clifton}.
\newblock {CLOCS: Contrastive Learning of Cardiac Signals Across Space, Time, and Patients}.
\newblock \emph{arXiv e-prints}, art. arXiv:2005.13249, May 2020.
\newblock \doi{10.48550/arXiv.2005.13249}.

\bibitem[Kolk et~al.(2024)Kolk, Ruip{\'e}rez-Campillo, Alvarez-Florez, Deb, Bekkers, Allaart, {Van Der Lingen}, Clopton, I{\v s}gum, Wilde, Knops, Narayan, and Tjong]{Kolk:2024:vae_ecg}
Maarten~Z.H. Kolk, Samuel Ruip{\'e}rez-Campillo, Laura Alvarez-Florez, Brototo Deb, Erik~J. Bekkers, Cornelis~P. Allaart, Anne-Lotte~C.J. {Van Der Lingen}, Paul Clopton, Ivana I{\v s}gum, Arthur~A.M. Wilde, Reinoud~E. Knops, Sanjiv~M. Narayan, and Fleur~V.Y. Tjong.
\newblock Dynamic prediction of malignant ventricular arrhythmias using neural networks in patients with an implantable cardioverter-defibrillator.
\newblock \emph{eBioMedicine}, 99:\penalty0 104937, 2024.
\newblock ISSN 2352-3964.
\newblock \doi{https://doi.org/10.1016/j.ebiom.2023.104937}.
\newblock URL \url{https://www.sciencedirect.com/science/article/pii/S2352396423005030}.

\bibitem[LeCun et~al.(1989)LeCun, Boser, Denker, Henderson, Howard, Hubbard, and Jackel]{LeCun:1989:cnn_original}
Y.~LeCun, B.~Boser, J.~S. Denker, D.~Henderson, R.~E. Howard, W.~Hubbard, and L.~D. Jackel.
\newblock Backpropagation applied to handwritten zip code recognition.
\newblock \emph{Neural Computation}, 1\penalty0 (4):\penalty0 541--551, 1989.
\newblock \doi{10.1162/neco.1989.1.4.541}.

\bibitem[{Loshchilov} and {Hutter}(2016)]{Loshchilov:2016:cosine_learning}
Ilya {Loshchilov} and Frank {Hutter}.
\newblock {SGDR: Stochastic Gradient Descent with Warm Restarts}.
\newblock \emph{arXiv e-prints}, art. arXiv:1608.03983, August 2016.
\newblock \doi{10.48550/arXiv.1608.03983}.

\bibitem[{Mehari} and {Strodthoff}(2021)]{Mehari:2021:ssl_for_12_lead}
Temesgen {Mehari} and Nils {Strodthoff}.
\newblock {Self-supervised representation learning from 12-lead ECG data}.
\newblock \emph{arXiv e-prints}, art. arXiv:2103.12676, March 2021.
\newblock \doi{10.48550/arXiv.2103.12676}.

\bibitem[Paszke et~al.(2019)Paszke, Gross, Massa, Lerer, Bradbury, Chanan, Killeen, Lin, Gimelshein, Antiga, Desmaison, K{\"o}pf, Yang, DeVito, Raison, Tejani, Chilamkurthy, Steiner, Fang, Bai, and Chintala]{paszke:2019:pytorch}
Adam Paszke, Sam Gross, Francisco Massa, Adam Lerer, James Bradbury, Gregory Chanan, Trevor Killeen, Zeming Lin, Natalia Gimelshein, Luca Antiga, Alban Desmaison, Andreas K{\"o}pf, Edward Yang, Zachary DeVito, Martin Raison, Alykhan Tejani, Sasank Chilamkurthy, Benoit Steiner, Lu~Fang, Junjie Bai, and Soumith Chintala.
\newblock Pytorch: An imperative style, high-performance deep learning library.
\newblock In \emph{Advances in Neural Information Processing Systems 32}, pages 8024--8035, 2019.

\bibitem[Ribeiro et~al.(2020)Ribeiro, Ribeiro, Paix{\~a}o, Oliveira, Gomes, Canazart, Ferreira, Andersson, Macfarlane, Meira~Jr., Sch{\"o}n, and Ribeiro]{Ribeiro:2020:ecg_resnet}
Ant{\^o}nio~H. Ribeiro, Manoel~Horta Ribeiro, Gabriela M.~M. Paix{\~a}o, Derick~M. Oliveira, Paulo~R. Gomes, J{\'e}ssica~A. Canazart, Milton P.~S. Ferreira, Carl~R. Andersson, Peter~W. Macfarlane, Wagner Meira~Jr., Thomas~B. Sch{\"o}n, and Antonio Luiz~P. Ribeiro.
\newblock Automatic diagnosis of the 12-lead ecg using a deep neural network.
\newblock \emph{Nature Communications}, 11\penalty0 (1):\penalty0 1760, 2020.
\newblock \doi{10.1038/s41467-020-15432-4}.
\newblock URL \url{https://doi.org/10.1038/s41467-020-15432-4}.

\bibitem[{Smith} et~al.(2021){Smith}, {Dherin}, {Barrett}, and {De}]{Smith:2021:implicit_regularization_sgd}
Samuel~L. {Smith}, Benoit {Dherin}, David G.~T. {Barrett}, and Soham {De}.
\newblock {On the Origin of Implicit Regularization in Stochastic Gradient Descent}.
\newblock \emph{arXiv e-prints}, art. arXiv:2101.12176, January 2021.
\newblock \doi{10.48550/arXiv.2101.12176}.

\bibitem[Venkatachalam et~al.(2011)Venkatachalam, Herbrandson, and Asirvatham]{Venkatachalam:2011:ecg_filtering}
K.L. Venkatachalam, J.E. Herbrandson, and S.J. Asirvatham.
\newblock Signals and signal processing for the electrophysiologist: part i: electrogram acquisition.
\newblock \emph{Circulation: Arrhythmia and Electrophysiology}, 4\penalty0 (6):\penalty0 965--973, December 2011.
\newblock \doi{10.1161/CIRCEP.111.964304}.

\bibitem[{Y{\`e}che} et~al.(2021){Y{\`e}che}, {Dresdner}, {Locatello}, {H{\"u}ser}, and {R{\"a}tsch}]{Hugo:2021:neighborhood_contrastive_learning}
Hugo {Y{\`e}che}, Gideon {Dresdner}, Francesco {Locatello}, Matthias {H{\"u}ser}, and Gunnar {R{\"a}tsch}.
\newblock {Neighborhood Contrastive Learning Applied to Online Patient Monitoring}.
\newblock \emph{arXiv e-prints}, art. arXiv:2106.05142, June 2021.
\newblock \doi{10.48550/arXiv.2106.05142}.

\end{thebibliography}

\newpage
\appendix

\section{ECG visualisation}
\label{app:ecg_visualisation}
\begin{figure}
\centering
\includegraphics[width=1.0\textwidth]{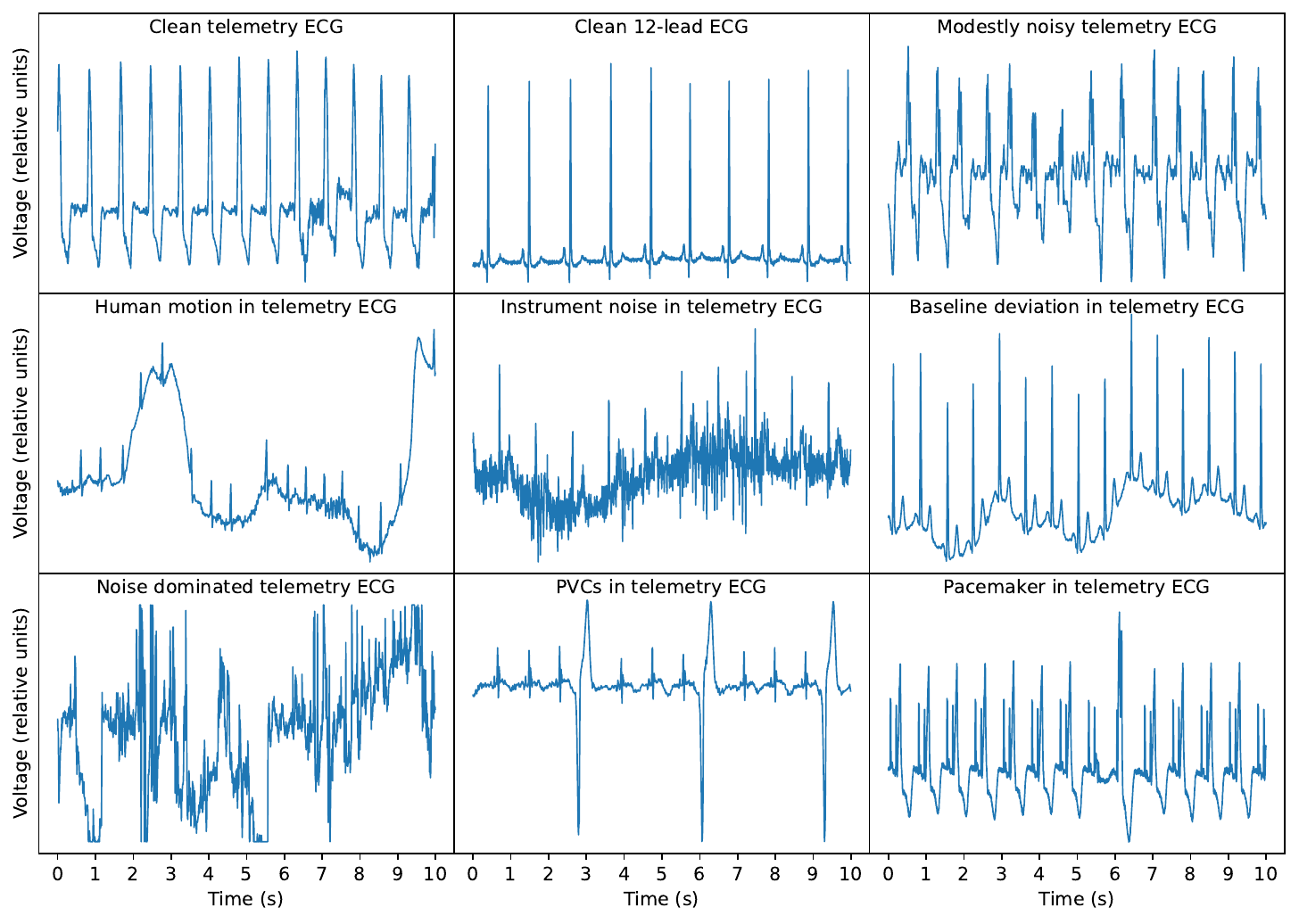}
\caption{Telemetry ECG segments can exhibit much more varied and noisy behavior than the average 12-lead ECG. Some features are physiological and can't be filtered or discarded.}
\label{fig:ecg_visualisation}
\end{figure}
\begin{figure}
\centering
\includegraphics[width=1.0\textwidth]{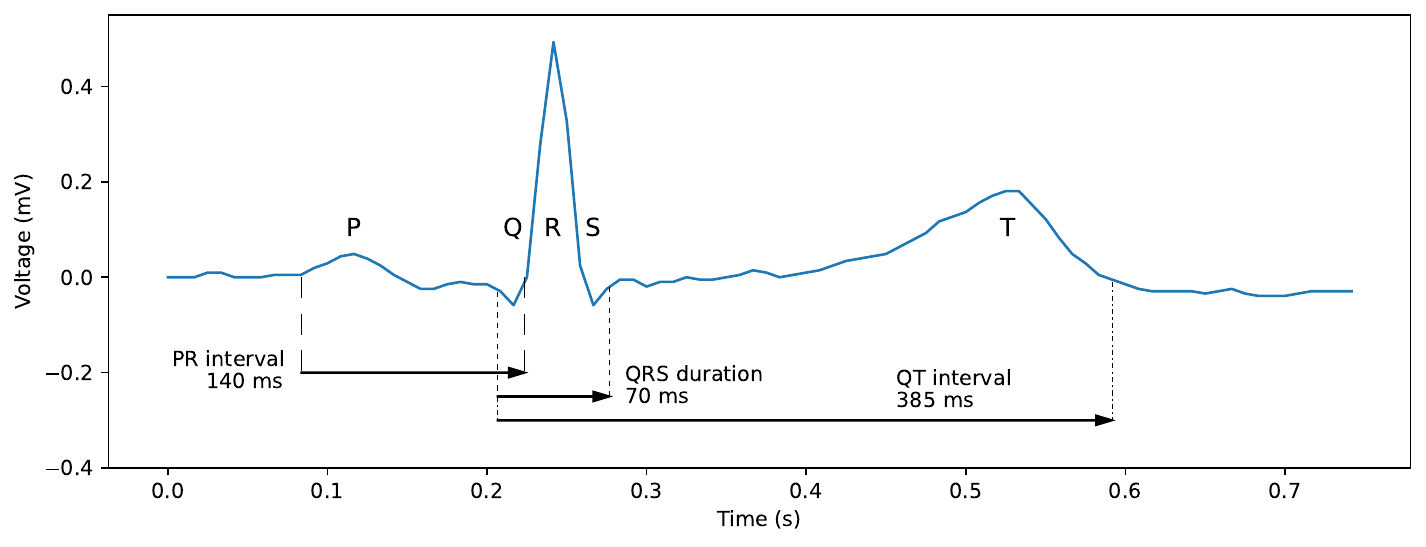}
\caption{The features P, Q, R, S and T arise from different key moments in the cardiac electrical cycle, and timing intervals between them carry important medical information. Features can very greatly in size, duration, and sign between different leads or physiological states. The ECG shown is from telemetry data recorded at 120Hz.}
\label{fig:pqrst_illustration}
\end{figure}
A sample set of various quality ECG signals are shown in Fig.~\ref{fig:ecg_visualisation}. The first two panels show qualitatively clean signals from telemetry monitoring and 12-lead ECG respectively. All other panels illustrate different types of noise seen in the telemetry data. Note that the voltage range is omitted in each panel for visualisation purposes, but each signal can vary greatly, with QRS peaks often surpassing $\sim 2$mV while other waveforms are completely within the $\pm0.4$mV range. The final two panels illustrate spikes which are not noise but rather the clinically relevant phenomena of premature ventricular contractions (PVCs, bottom-center) and pacemaker signals (bottom-right), showing anomalous behavior which, unlike noise, a neural network may need to interpret for medical tasks. The final two examples also illustrate the difficulty of a simple Fourier based filtering or selection of signals, since these physiological and spatially localized features in the waveform extend to all frequencies in Fourier space.

The morphology of a typical heartbeat is shown in Fig.~\ref{fig:pqrst_illustration} together with annotated timings which are relevant for the intervals downstream task.

\section{Encoder architecture}
\label{app:architecture}
The model architecture used throughout this paper is based on ResNet \cite{He:2015:resnet_original, He:2016:resnet_v2}, however with some adaptations for 1D signals. Most obviously all convolution \cite{LeCun:1989:cnn_original} and batchnorm \cite{Ioffe:2015:batchnorm_original} layers are 1D instead of the original 2D. The first convolution layer ingests a signal 1024 indices long ($\sim 8.5$s) with four channels corresponding to the four ECG leads (generally leads I, II, III and V1), and outputs a signal half the length with {\tt chan\_start} channels, which is a hyperparameter. Following \cite{He:2015:resnet_original} we compose four blocks with a variable number of layers connected by skip connections, such that the final output channels are {\tt chan\_out}$=2^3\times${\tt chan\_start}, with the length similarly divided. We fix the initial convolution kernel size to be 15, and all other kernel sizes to be 3 (based on initial testing). Additionally the local max pooling layers are replaced with average pools, as these yielded better performance on the intervals task (tested via supervised learning). We set the number of layers in each block to match \cite{He:2015:resnet_original} but set the channel numbers to more sensible values for 1D signals. In Table~\ref{tab:model_specs} we give fiducial models with corresponding model sizes. Note that 50, 101 and 152 are bottleneck networks, and with the smaller channel sizes this means the number of parameters do not grow monotonically with layer count.
\begin{table}[ht]
\centering
\caption{Model specifications and trainable parameters for 1D ResNets without projection heads}
\label{tab:model_specs}
\begin{tabular}{cccc}
\toprule
Model name & Chan start & Chan out & Trainable params (M) \\
\midrule
ResNet18 & 16 & 128 & 0.24\\
ResNet34 & 32 & 256 & 1.81\\
ResNet50 & 32 & 256 & 0.26\\
ResNet101 & 32 & 256 & 0.45\\
ResNet152 & 64 & 512 & 2.43\\
\vspace{-0.2cm}\\
ResNet18x2 & 32 & 256 & 0.96\\
ResNet34x2 & 64 & 512 & 7.22\\
ResNet50x2 & 64 & 512 & 1.01\\
ResNet101x2 & 64 & 512 & 1.79\\
ResNet152x2 & 128 & 1024 & 9.64\\
\bottomrule
\end{tabular}
\end{table}

\section{Training hyperparameters}
\label{app:hyperparameters}
\begin{table}[ht]
\centering
\caption{All hyperparameters for the pretraining stage}
\label{tab:pretraining-hyperparameters}
\begin{tabular}{cccccc}
\toprule
Parameter & Value & \hspace{0.5cm} & Parameter & Value \\
\midrule
Queue size & 38912 & & Total epochs & 500 \\
Temperature & 0.1 & & Final learning rate & $10^{-6}$ \\
Batch size & 256 & & Validation batch size & 512 \\
Initial learning rate & 0.000625 & & Momentum & 0.999 \\
Warmup epochs & 5 & & Weight decay & 0.0001 \\
\bottomrule
\end{tabular}
\end{table}
The hyperparameters used for the pretraining stage are provided in Table~\ref{tab:pretraining-hyperparameters}. The model is trained using a CosineAnnealing learning rate schedule \cite{Loshchilov:2016:cosine_learning} with a linear warmup schedule \cite{he:2018:bag_of_tricks} starting from $1/10$ of the learning rate. The Adam optimizer is employed with default parameters \cite{Kingma:2014:adam_optimizer}.

\begin{table}
\centering
\caption{All hyperparameters for training on each downstream task}
\begin{tabular}{ccccccc}
\toprule
Task & Type & Batch & Epochs & Learning rate & Scheduler step & Warmup \\ \midrule
\multirow{3}{*}{Age \& sex} & Supervised & 512 & 60 & $5\times 10^{-3}$ & 20 &  5 \\
                     & Linear Head & 1024 & 15 & $10^{-1}$ & 5 & 0 \\
                     & Fint-tuning & 1024 & 30 & $10^{-4}$ & 20 & 0 \\ \midrule
\multirow{3}{*}{Intervals} & Supervised & 256 & 60 & $5\times 10^{-3}$ & 20 & 5 \\
                     & Linear Head & 512 & 15 & $2\times 10^{-1}$ & 5 & 0 \\ 
                     & Fint-tuning & 512 & 45 & $2\times 10^{-4}$ & 20 & 0 \\ \midrule
\multirow{3}{*}{AFib} & Supervised & 128 & 60 & $2\times 10^{-3}$ & 20 & 5 \\ 
                     & Linear Head & 128 & 15 & $10^{-1}$ & 5 & 0 \\ 
                     & Fint-tuning & 128 & 50 & $10^{-4}$ & 20 & 0 \\ \bottomrule
\end{tabular}
\label{table:downstream_hyperparameters}
\end{table}
The hyperparameters for downstream tasks are provided in Table~\ref{table:downstream_hyperparameters}. The fine-tuning stage always follows training the linear head. The optimizer for all tasks is Adam with default parameters \cite{Kingma:2014:adam_optimizer}, and the step scheduler multiples the learning rate by $0.1$ at the milestone indicated in the table. For each task the epoch with highest validation figure of merit is kept (MAE for age, AUC for sex/Afib, MAPE for intervals). For ResNet152 and ResNet152x2 the learning rate was halved from the usual value, and for ResNet152x2 the batch size was reduced to fit in VRAM while using gradient accumulation to compensate for the change.

\section{Study comparison}
\label{app:study_comparison}
\begin{table}
\centering
\caption{Summary of figures of merit for each downstream task across different studies}
\label{table:simplified_figures_of_merit}
\begin{tabular}{ccccc}
\toprule
Task & Metric & Result Range & Sources & This work \\
\midrule
\multirow{2}{*}{Age regression (years)} & MAE & 6.90\,--\,8.22 & \cite{Zachi:2019:ecg_age_sex}, \cite{Abbaspourazad:2023:foundation_wearable} & 8.07 \\
 & $R^2$ & 0.60\,--\,0.70 & \cite{Zachi:2019:ecg_age_sex}, \cite{Diamant:2022:pclr} & 0.64\\
 \vspace{-0.2cm}\\
\multirow{2}{*}{Sex classification} & AUC & 0.88\,--\,0.968 & \cite{Zachi:2019:ecg_age_sex}, \cite{Abbaspourazad:2023:foundation_wearable} & 0.88 \\
 & $F1$ & 0.84\,--\,0.87 & \cite{Zachi:2019:ecg_age_sex}, \cite{Diamant:2022:pclr} & 0.83 \\
 \vspace{-0.2cm}\\
\multirow{2}{*}{QT interval regression (ms)} & MAE & 11.2\,--\,12.3 & \cite{Alam:2024:qt_interval} & 13.1 \\
 & $R^2$ & 0.87\,--\,0.92 & \cite{Alam:2024:qt_interval} & 0.83 \\
 \vspace{-0.2cm}\\
\multirow{2}{*}{HR regression (bpm)} & MAE & 1.08\,--\,1.42 & \cite{Alam:2024:qt_interval} & 1.67 \\
 & $R^2$ & 0.96\,--\,0.99 & \cite{Alam:2024:qt_interval} & 0.95 \\
 \vspace{-0.2cm}\\
 \multirow{2}{*}{Afib classification} & AUC & 0.85\,--\,0.97 & \cite{Zachi:2019:ecg_resnet_afib_original}, \cite{Hannun:2019:ambulatory_afib} & 0.99 \\
 & $F1$ & 0.70\,--\,0.84 & \cite{Diamant:2022:pclr},\cite{Hannun:2019:ambulatory_afib} & 0.93 \\
\bottomrule
\end{tabular}
\end{table}
\begin{table}
\centering
\caption{Comparison of relevant studies using ECG data}
\label{table:study_comparison}
\begin{tabular}{ccccc}
\toprule
Study & \# patients & \# ECGs & ECG type & Context\\ \midrule
\cite{Zachi:2019:ecg_age_sex} & 770k & 770k & 10s, 12-lead & Supervised learning \\
\cite{Diamant:2022:pclr} & 404k & 3.2M & 10s, 12-lead  & PCLR SSL \\
\cite{Abbaspourazad:2023:foundation_wearable} & 106k & 3.7M & 30s, 1-lead  & Apple watch, SSL \\
\cite{Alam:2024:qt_interval} & 903k & 4.22M & 10s, 1-lead & Supervised learning \\
\cite{Zachi:2019:ecg_resnet_afib_original} & 181k & 650k & 10s, 12-lead & Supervised learning \\
\cite{Hannun:2019:ambulatory_afib} & 54k & 91k & 1-lead & Supervised learning \\
This work & 72k & 8.8M & 60s, 4-lead & Telemetry monitoring, PCLR SSL \\
\bottomrule
\end{tabular}
\end{table}
In Table~\ref{table:simplified_figures_of_merit} the main figures of merit are summarised: MAE, MAPE and $R^2$ for regression tasks, AUC and $F1$ for classification tasks. Values from this work are provided for each of the metrics using ResNet33x2, however we note that the time of downstream fine-tuning we select for the epoch with highest validation MAE (age), AUC (sex/Afib) and MAPE (intervals) -- selecting the optimal epochs for other metrics should yield very similar results. In this work the four intervals are minimised simultaneously, but results in Table~\ref{table:study_comparison} are provided on individual tasks to compare with \cite{Alam:2024:qt_interval}. Here again performance could potentially improve with a dedicated network for each interval, but early testing suggested the differences are negligible. The results taken from \cite{Abbaspourazad:2023:foundation_wearable} are those pretrained at patient level and evaluated at segment level, i.e. not those taken from averaging embeddings over patients.

In Table~\ref{table:study_comparison} the various demographics and ECG signal origins for different studies are summarised. The numbers provided for this work correspond to the pretraining dataset. In particular the intervals and Afib downstream tasks had far fewer labels. While the dataset used in this work contains 60s ECG segments only $\sim 8.5$s are ingested by the model at a time which is chosen randomly at the time of data loading, thus providing one source of randomness between epochs.

\section{Further downstream results}
\label{app:absolute_downstream}
\begin{figure}
\centering
\includegraphics[width=1.0\textwidth]{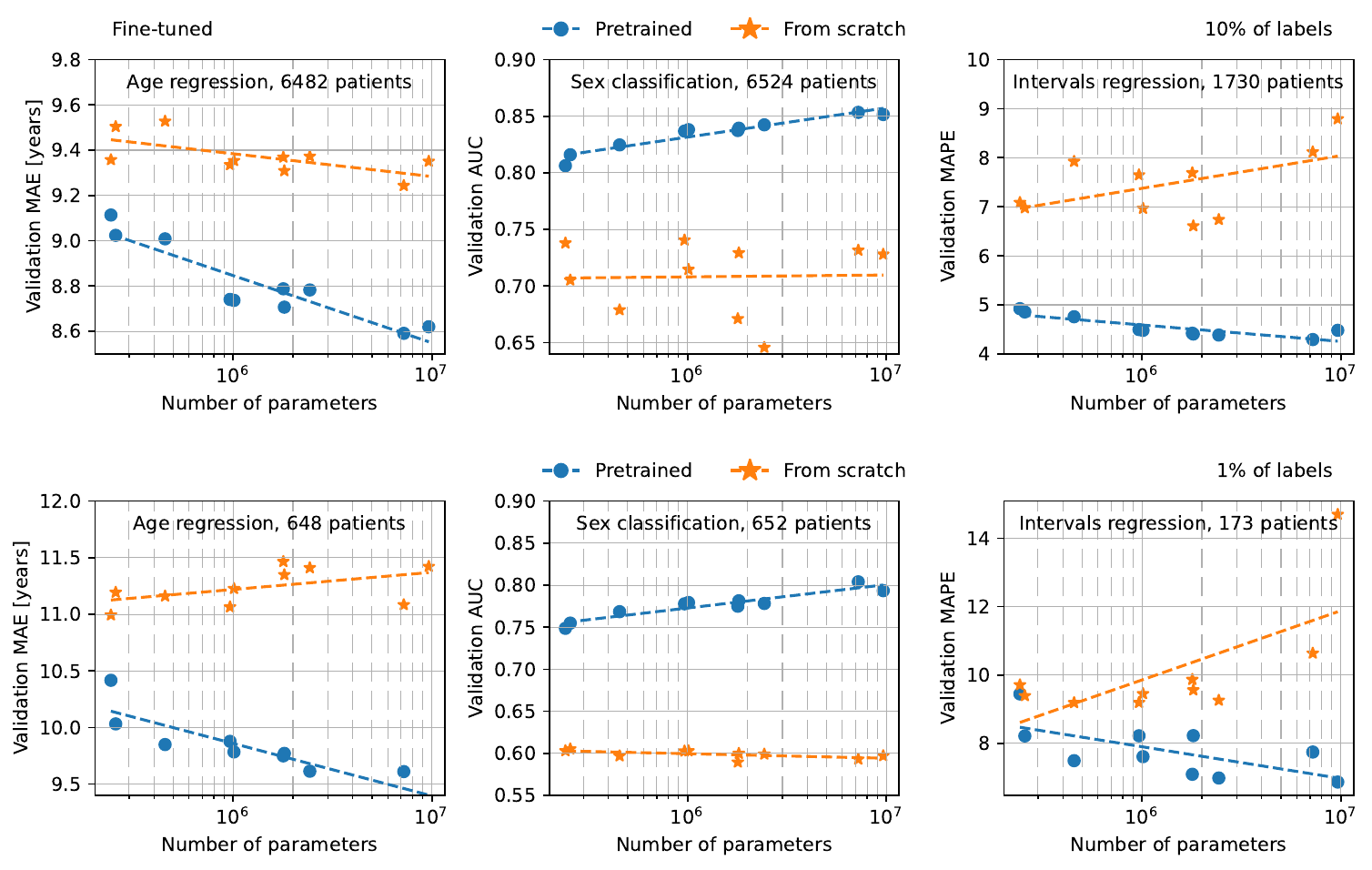}
\caption{Pretrained and fine-tuned models show a greater advantage over supervised models with less available labels. With $1\%$ of labels (bottom row) the supervised models typically perform worse with model size, while large pretrained models still find better minima. The results for 100\% of labels are shown in Fig.~\ref{fig:proxy_downstream_absolute_100}.}
\label{fig:proxy_downstream_absolute_grid}
\end{figure}
\begin{figure}
\centering
\includegraphics[width=1.0\textwidth]{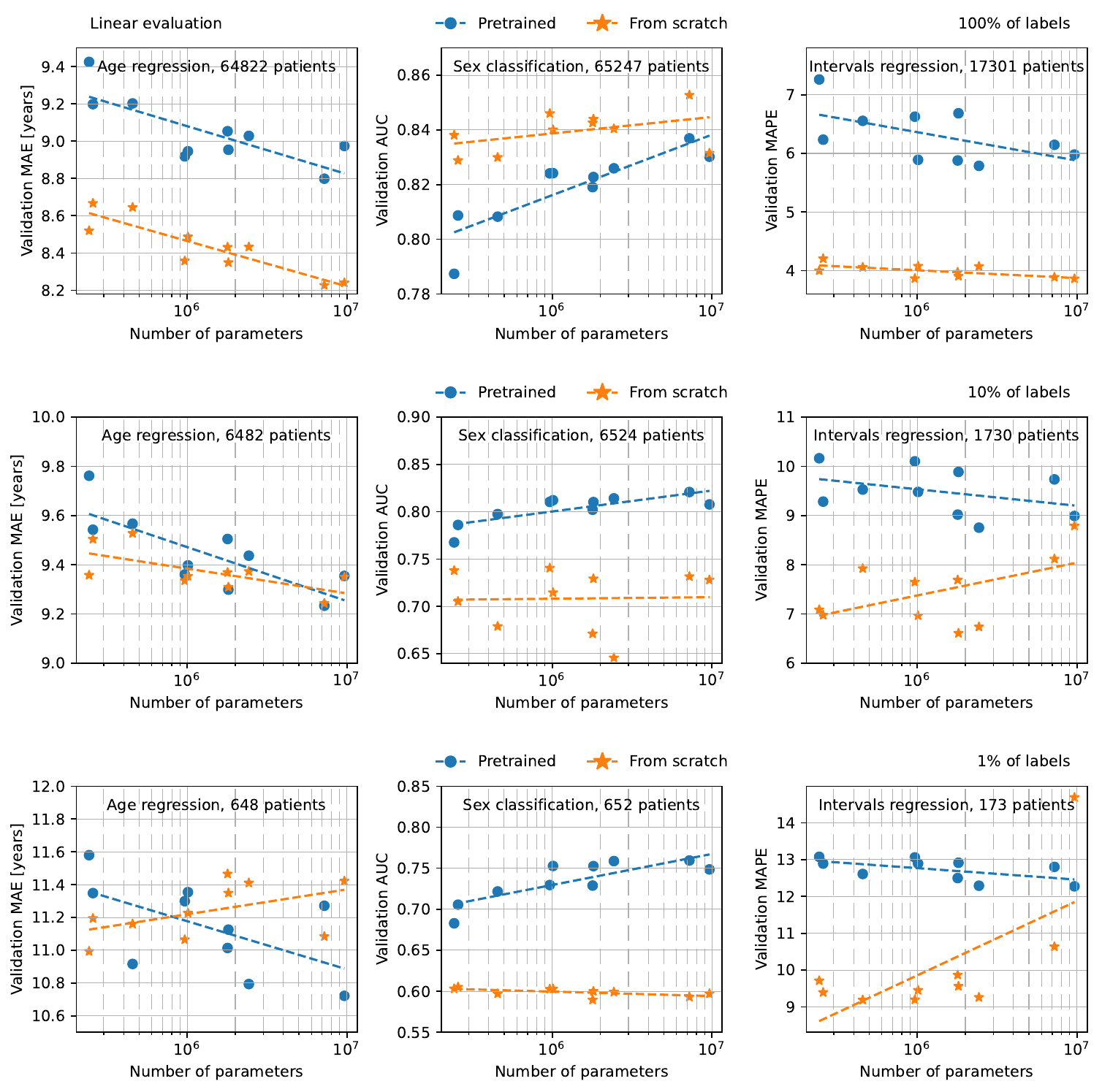}
\caption{Large pretrained models which undergo a linear evaluation with frozen weights still outperform supervised models for scarce labels on the age (left column) and sex (middle column) tasks. Fine-tuning is evidently necessary to use the extra information in large label amounts, as evidenced by supervised models outperforming pretrained ones for $100\%$ of labels (top row) but not for 10\% (middle row) or 1\% (bottom row). The intervals task (right column) does not favor any pretrained model without fine-tuning, showing a potential inductive bias in pretraining.}
\label{fig:proxy_downstream_linear_eval_absolute_grid}
\end{figure}
Fig.~\ref{fig:proxy_downstream_absolute_grid} provides the downstream results for 10\% and 1\% of labels in terms of absolute figures of merit (not relative improvement). It is noteworthy that in many cases a \% improvement with model size was due \textbf{both} to pretrained models performing better as well as supervised models performing worse.
Similar results without fine-tuning -- just linear evaluation on frozen embeddings -- are shown in Fig.~\ref{fig:proxy_downstream_linear_eval_absolute_grid}. Comparing to supervised models trained on large label numbers renders the pretrained models inefficient. However, when reducing the label amount the pretrained advantage returns, at least for age and sex. For the intervals task the pretrained model never matches the supervised model, even for $1\%$ of labels. This again hints towards an inductive bias in the pretraining that can be somewhat overcome with fine tuning, but not with frozen embeddings.

\end{document}